\documentstyle[12pt]{article}
\begin{document}

{\Large \bf ON POLINOMIAL FORMULATIONS OF GENERAL RELATIVITY}\\
{\bf Vladimir Khatsymovsky}\footnote{Budker Institute of Nuclear
Physics, Novosibirsk 630090, Russia}\\
----------------------------------------------------------------------
--\\
As basic variables in general relativity (GR) are chosen
antisymmetric
connection and bivectors - bilinear in tetrad area tensors subject
to appropriate (bilinear) constraints. In canonical formalism we
get theory with polinomial constraints some of which are II class.
On partial resolving the latter we get another polinomial
formulation.
Separating self- and antiselfdual parts of antisymmetric tensors we
come to Ashtekar constraints including those known as "reality
conditions" which connect self- and antiselfdual sectors of the
theory.
These conditions form second class system and cannot be simply
imposed
on quantum states (or taken as initial conditions in classical
theory).
Rather these should be taken into account in operator sence by
forming
corresponding Dirac brackets. As a result, commutators between
canonical
variables are no longer polinomial, and even separate treatment of
self-
and antiselfdual sectors is impossible.\\
----------------------------------------------------------------------
----\\
\newpage
1. Ashtekar variables \cite{Ash} attract much attention as possible
tool to solve quantum constraints of GR nonperturbatively
\cite{Jac}.
Such a possibility is connected with polinomiality of GR in the new
variables. In this note two another polinomial versions of GR are
suggested.
Canonical formalism is developed and connection with Ashtekar
variables
is considered.

The issue point is Einstein-Hilbert action in the tetrad-connection
variables \cite{Schw}:
\begin{equation}
S=\frac{1}{8}\int \! d^{4} x \: \epsilon_{abcd} \epsilon^{\mu\nu
\lambda\rho} e^{a}_{\mu} e^{b}_{\nu} [{\cal D}_{\lambda},{\cal D}
_{\rho}]^{cd}
\end{equation}

\noindent where ${\cal
D}_{\lambda}=\partial_{\lambda}+\omega_{\lambda}$
(in fundamental representation) is covariant derivative, and
$\omega_{\mu}^{ab}=-\omega_{\mu}^{ba}$ is element of $so(3,1)$, Lie
algebra
of $SO(3,1)$ group. Raising and lowering indices is performed with
the help
of metric $\eta_{ab}={\rm diag}(-1,1,1,1)$, while
$\epsilon^{0123}=+1$.
$\alpha,\beta, \ldots =1,2,3$ and $\mu,\nu, \ldots =0,1,2,3$ are
coordinate
indices and $a,b, \ldots =0,1,2,3$ are local ones. Separating space
and
time indices we put Lagrangian density into the form
\begin{equation}
{\cal L}_{0}=\pi^{\alpha} \circ \dot{\omega}_{\alpha}-h \circ {\cal
D}
_{\alpha}\pi^{\alpha} - n_{\alpha} \circ R^{\alpha}
\label{Lagr0}
\end{equation}

\noindent Here
\begin{equation}
h=-\omega_{0},~~~R_{ab}^{\alpha}=\frac{1}{2}\epsilon
^{\alpha\beta\gamma}[{\cal D}_{\beta},{\cal D}_{\gamma}]_{ab},~~~
\pi_{ab}^{\alpha}=\frac{1}{2}\epsilon_{abcd}\epsilon^{\alpha\beta
\gamma}e_{\beta}^{c}e_{\gamma}^{d},~~~n_{\alpha ab}=-\epsilon_{abcd}
e_{0}^{c}e_{\alpha}^{d}
\label{def-biv}
\end{equation}
Scalar product of two matrices $(\circ)$ and hereafter used their
dual
product $(*)$ and dual matrix are defined as
\begin{eqnarray}
A \circ B & \stackrel{\rm def}{=} &
\frac{1}{2}A^{ab}B_{ab}\nonumber\\
A*B & \stackrel{\rm def}{=} & A \circ (\,^{*}\!B)\\
^{*}\!B^{ab} & \stackrel{\rm def}{=} &
\frac{1}{2}\epsilon^{ab}_{~~cd}
B^{cd}\nonumber
\end{eqnarray}

In order that $n_{\alpha}, \pi^{\alpha}$ be of the form pointed out
in (\ref{def-biv}), dual products of the type $\pi * \pi, n*n$ and
traceless part of $\pi * n$ should vanish:
\begin{eqnarray}
\pi^{\alpha}*\pi^{\beta} & = & 0
\label{pi-pi-cond}\\
\pi^{\alpha}*n_{\beta} & - &\frac{1}{3}(\pi^{\gamma}*n_{\gamma})
\delta^{\alpha}_{\beta}=0
\label{pi-n-cond}\\
n_{\alpha}*n_{\beta} & = & 0
\label{nn-cond}
\end{eqnarray}

\noindent Conversely, having got 6 antisymmetric matrices
$\pi^{\alpha},
n_{\alpha}$ (36 components), subject to 20 conditions
(\ref{pi-pi-cond})-
(\ref{nn-cond}), one can check that there exists the unique, up to
an
overall sign, tetrad $e^{a}_{\mu}$ ( $16=36-20$ components),in terms
of
which $\pi^{\alpha},n_{\alpha}$ are expressible according to
(\ref{def-biv}).
Adding (\ref{pi-pi-cond}) - (\ref{nn-cond}) to (\ref{Lagr0}) with the
help
of Lagrange multipliers gives
\begin{eqnarray}
{\cal L} & = & {\cal L}_{0}-\frac{1}{2}\mu_{\alpha\beta}
\pi^{\alpha}*\pi^{\beta}-\lambda^{\beta}_{\alpha}\pi^{\alpha}
*n_{\beta}-\frac{1}{2}\nu^{\alpha\beta}n_{\alpha}*n_{\beta}\\
 & \stackrel{\rm def}{=} & {\cal L}_{0}-\Lambda\Phi,\nonumber
\end{eqnarray}

\noindent where $\mu_{\alpha\beta},\nu^{\alpha\beta}$ are
symmetrical,
${\rm
tr}\lambda=\lambda^{\alpha}_{\alpha}=0,~\Lambda=(\mu,\nu,\lambda)$,
and $\Phi$ denotes the set of constraints (\ref{pi-pi-cond}) -
(\ref{nn-cond}).

\bigskip
2. Consider the structure of constraints. In the Lagrangian formalism
we
first vary ${\cal L}$ in $\Lambda,h,n$. This gives, together with
earlier introduced $\Phi$, also Gauss law
\begin{equation}
C\stackrel{\rm def}{=}{\cal D}_{\alpha}\pi^{\alpha}=0
\end{equation}

\noindent and
\begin{equation}
\lambda_{\beta}^{\alpha}\,^{*}\!\pi^{\beta}
+\nu^{\alpha\beta}\,^{*}\!n_{\beta}+R^{\alpha}=0
\label{lambda-nu-eq}
\end{equation}

\noindent Let us multiply (\ref{lambda-nu-eq}) in scalar way by
$\pi^{\gamma},
n_{\gamma}$. This allows, with the help of $\Phi$, to find
$\lambda,\nu$;
besides, requiring $\nu^{\alpha\beta}$ be symmetrical gives
constraints,
having the form of momentum ones \cite{Ash} - combinations of
diffeomorphism
generators with local rotation ones $C$,
\begin{equation}
{\cal H}_{\alpha}\stackrel{\rm def}{=}\epsilon_{\alpha\beta\gamma}
\pi^{\beta}\circ R^{\gamma}=0,
\label{H-alpha}
\end{equation}

\noindent while requirement ${\rm tr}\lambda=0$ leads to Hamiltonian
constraint\footnote{More accurately, our constraint ${\cal H}_{0}$
is
combination of Hamiltonian constraint \cite{Ash}, whose effect in
combination with $C$ are shifts in time, and of ${\cal H}_{\alpha}$;
see below}
\begin{equation}
{\cal H}_{0}\stackrel{\rm def}{=}n_{\alpha}\circ R^{\alpha}=0
\label{H0}
\end{equation}

\noindent Constraints (\ref{H-alpha}) and (\ref{H0}) are produced by
varying action by operators
\begin{equation}
n_{\alpha} \circ \frac{\delta}{\delta n_{\alpha}}
\end{equation}

\noindent and
\begin{equation}
\epsilon_{\alpha\beta\gamma}\pi^{\beta} \circ \frac{\delta}
{\delta n_{\gamma}}
\end{equation}

Upon varying w.r.t. $\pi,\omega$ we get equations of motion
\begin{eqnarray}
\dot{\omega}_{\alpha} & = &-{\cal D}_{\alpha}h+\,^{*}\!(\mu\pi+
\lambda n)_{\alpha}
\label{omega-dot}\\
\dot{\pi}^{\alpha} & = & [h,\pi^{\alpha}]-
\epsilon^{\alpha\beta\gamma}{\cal D}_{\beta}n_{\gamma}
\label{pi-dot}
\end{eqnarray}

Further require that constraints obtained be conserved in time. Eqs.
(\ref{pi-n-cond}) and (\ref{nn-cond}) (14 components) allow to find
18 components $n_{\alpha}$ up to 4 parameters. Differentiating these
will give $\dot{n}_{\alpha}$ with the same degree of undefiniteness.
Namely,
\begin{equation}
\dot{n}_{\alpha}=vn_{\alpha}+\epsilon_{\alpha\beta\gamma}
u^{\beta}\pi^{\gamma},
\label{n-dot}
\end{equation}

\noindent where $u^{\alpha},v$ are parameters. Knowing
$\dot{n}_{\alpha}$
we can differentiate ${\cal H}_{0}$ and parameters $\lambda,\nu$
earlier
obtained from (\ref{lambda-nu-eq}). The rest of constraints can be
differentiated with the help of (\ref{omega-dot}) and (\ref{pi-dot})
without
problem. The Gauss law, Hamiltonian and momentum constraints are
conserved
identically, and the only nontrivial is condition
\begin{equation}
\frac{d}{dt}(\pi^\alpha * \pi^\beta)=0,
\end{equation}

\noindent which gives the constraint
\begin{equation}
G^{\gamma\delta}\stackrel{\rm def}{=}n_{\alpha}*
(\epsilon^{\alpha\beta\gamma}{\cal D}_{\beta}\pi^{\delta}+
\epsilon^{\alpha\beta\delta}{\cal D}_{\beta}\pi^{\gamma})=0.
\label{n-D-pi}
\end{equation}

\noindent It is thus the consequence of equation of motion for
connection
$\delta S/\delta\omega=0$.

Finally, differentiating (\ref{n-D-pi}) allows us to find
$\mu_{\alpha
\beta}$. Indeed, dependence on $\mu_{\alpha\beta}$ arises due to
terms
with $\dot{\omega}$, see (\ref{omega-dot}), and has the form
\begin{equation}
\dot{G}^{\alpha\beta}=2\det\| g_{\alpha\beta}\|\epsilon_{abcd}e^{a}
_{0}e^{b}_{1}e^{c}_{2}e^{d}_{3}(g^{\gamma\delta}g^{\epsilon\zeta}-
g^{\gamma\epsilon}g^{\delta\zeta})\mu_{\epsilon\zeta}+\cdots,\\
{}~~~g_{\alpha\beta}\stackrel{\rm def}{=}e^{a}_{\alpha}e_{\beta a}.
\end{equation}

\noindent Due to nondegeneracy of metric the equation $\dot{G}=0$
is uniquely solvable for $\mu$.

In Hamiltonian formalism denote by $\tilde{q}$ the momentum,
conjugate
to coordinate $q$. In particular, $(\pi,\omega)$ already form
canonical
pair $(\tilde{q},q)$. Hamiltonian density is
\begin{equation}
{\cal H}=\sum_{q}\tilde{q}\dot{q}-{\cal L}.
\end{equation}

\noindent First, the primary constraints can be found:
\begin{equation}
\tilde{\Lambda}=0,~~~\tilde{h}=0,~~~\tilde{n}=0.
\end{equation}

\noindent Their conservation leads to secondary constraints
\begin{equation}
0=\frac{d}{dt}\tilde{q}=\{\tilde{q},\int\!{\cal H}\,d^{3}x\}=
-\frac{\partial{\cal H}}{\partial q},~~~(q\neq\pi,\omega),
\end{equation}

\noindent which are easily recognised to be the earlier obtained in
Lagrangian formalism constraints $\Phi,~C$ and (\ref{lambda-nu-eq}).
The Poisson bracket is defined as usual, in particular
\begin{equation}
\{\omega_{\alpha}^{ab}(x),\pi_{cd}^{\beta}(x^{\prime})\}
=(\delta^{a}_{c}\delta^{b}_{d}-\delta^{a}_{d}\delta^{b}_{c})
\delta^{\beta}_{\alpha}\delta^{3}(x-x^{\prime})
\end{equation}

\noindent The further Dirac procedure of extracting the constraints
completely repeats the above consideration in the Lagrangian
formalism.
As a result, the following nondynamical (i.e. different from
$\pi,\omega$) variables remain undefined:
$h,~\dot{h}$, being Lagrange multipliers at constraints $C,
{}~\tilde{h}$; 4 parameters in $n$ and the same number of those in
$\dot{n}$
(see (\ref{n-dot})) being Lagrange multipliers at constraints
${\cal H}_\mu$ and at four combinations of $\tilde{n}$,
respectively.
This means that corresponding constraints are I class.

In particular, to describe evolution of physical observables (which
are
natural to thought of as functions of $\pi,~n,~\omega$) it is
sufficient
to use the set of pairs $(\pi,\omega),(\tilde{n},n)$ as phase space.
Then phase manifold of GR is defined by I class constraints,
\begin{equation}
C,~~~{\cal H}_{\mu},~~~{\rm 4~combinations}~{\tilde{n}},
\label{I-class}
\end{equation}

\noindent and by the others, II class ones:
\begin{equation}
\Phi,~~~G^{\alpha\beta},~~~{\rm 14~combinations}~{\tilde{n}}.
\label{II-class}
\end{equation}

\noindent The number of the degrees of freedom is equal to the number
of
canonical pairs minus the number of I class constraints and half of
the number of II class ones. Let $[A]$ be the number of components of
a
value $A$. Then the number of I class combinations of ${\tilde{n}}$
(4 in
(\ref{I-class})) arises as $[n]-[\Lambda]+[\mu]$. The same is the
number of
constraints ${\cal H}_{\mu}$. As a result, the number of the degrees
of
freedom turns out to be expressible as
\begin{equation}
[\omega]-[n]-[h]+[\Lambda]-2[\mu]=2,
\end{equation}

\noindent as it would expected.

\bigskip
3. Some disadvantage of formulation (\ref{I-class}), (\ref{II-class})
is
that $\tilde{n}$ are not purely I or II class constraints but rather
their
nontrivial combinations. We can pass to another polinomial version of
GR by
noting that $\Phi$ can be solved for $n$ as
\begin{equation}
n=\epsilon_{\alpha\beta\gamma}w^{\beta}\pi^{\gamma}+
v\epsilon_{\alpha\beta\gamma}\pi^{\beta}\pi^{\gamma},
\end{equation}

\noindent where $w^{\alpha},v$ are parameters. Then constraints
$G^{\alpha\beta},~{\cal H}_{0}$ at $v\neq 0$ are equivalent to the
following ones:
\begin{eqnarray}
\tilde{G}^{\alpha\beta} & \stackrel{\rm def}{=} & \pi^{\gamma}*
([\pi^{\alpha},{\cal D}_{\gamma}\pi^{\beta}]+
([\pi^{\beta},{\cal D}_{\gamma}\pi^{\alpha}])\\
\tilde{\cal H}_{0} & \stackrel{\rm def}{=} & \epsilon_{\alpha\beta
\gamma}\pi^{\alpha}\pi^{\beta}\circ R^{\gamma}
\end{eqnarray}

\noindent Phase manifold in terms of $(\pi,\omega)$ is given by
constraints
\begin{equation}
C,~~~{\cal H}_{\alpha},~~~\tilde{\cal H}_{0}
\label{I-class-2}
\end{equation}

\noindent and
\begin{equation}
\pi^{\alpha}*\pi^{\beta},~~~\tilde{G}^{\alpha\beta}
\label{pi-omega-II-class}
\end{equation}

\noindent of I and II class, respectively.

In quantum theory II class constraints cannot be simply imposed on
states
since due to their noncommutativity this will lead to vanishing the
wavefunction itself. Instead, these should be taken into account in
the
operator sence by assigning to quantum commutators the values of the
Dirac
rather then Poisson brackets. Dirac brackets arise from Poisson ones
when
projecting orthogonally to the II class constraint surface in the
phase
space:
\begin{equation}
\{f,g\}_{\rm D}\stackrel{\rm def}{=}\{f,g\}-\{f,\Theta_{A}\}
(\Delta^{-1})^{AB}\{\Theta_{B},g\},
\end{equation}

\noindent where $\{\Theta_{A}\}$ is the full set of II class
constraints,
and $\Delta^{-1}$ is matrix inversed to that of their Poisson
brackets:
\begin{equation}
(\Delta^{-1})^{AB}\{\Theta_{B},\Theta_{C}\}=\delta^{A}_{C}
\end{equation}

Now when $\Theta_{A}$ are constraints (\ref{pi-omega-II-class}),
the matrix $\Delta^{-1}$ is easy to find. Poisson brackets on
constraint
surface take the form
\begin{eqnarray}
\{\int\!(\frac{1}{2}\mu_{\alpha\beta}\pi^{\alpha}*\pi^{\beta}
+\frac{1}{2}m_{\alpha\beta}\tilde{G}^{\alpha\beta})\,d^{3}x,
\int\!(\frac{1}{2}\mu^{\prime}_{\alpha\beta}\pi^{\alpha}*\pi^{\beta}
+\frac{1}{2}m^{\prime}_{\alpha\beta}\tilde{G}^{\alpha\beta})\,d^{3}x\}
=
\nonumber\\
\int\!(\det\| g_{\alpha\beta}\|)^{2}[{\rm tr}(m\mu^{\prime}-
\mu m^{\prime})+{\rm tr}\mu{\rm tr}m^{\prime}-
{\rm tr}m{\rm tr}\mu^{\prime}+~~~~~~~~~~\\
{\rm tr}(m\chi){\rm tr}m^{\prime}-
{\rm tr}m{\rm
tr}(m^{\prime}\chi)]\,d^{3}x,~~~~~~~~~~~~~~~~~~~~\nonumber\\
\chi_{\alpha\beta}\stackrel{\rm def}{=}\epsilon_{\alpha\gamma\delta}
\pi^{\gamma}*{\cal D}_{\beta}\pi^{\delta}.~~~~~~~~~~~~~~~~~~~~~~~~~
{}~~~~~~~~~~~~~~~~~~~~~~~~~~~~~~~~~~~\nonumber
\label{commut-II-class}
\end{eqnarray}

\noindent Here $m,m^{\prime},\mu,\mu^{\prime}$ are test functions
(symmetric matrices), while raising and lowering indices is made
with
the help of metric $g_{\alpha\beta}$. Inverting the bilinear form
(\ref{commut-II-class}) which leads to $\Delta^{-1}$ offers no
difficulties,
and Dirac bracket of any quantities $f,g$ turns out to be local:
\begin{eqnarray}
\lefteqn{\{f,g\}_{\rm D}=\{f,g\}-}\nonumber\\
 & & \frac{1}{2}\int\!(\det\| g_{\alpha\beta}\|)^{-2}
[{\rm tr}(\{f,\phi\}\{\tilde{G},g\}
-\{f,\tilde{G}\}\{\phi,g\})-\nonumber\\
\label{Dir-brack}
 & & \frac{1}{2}{\rm tr}\{f,\phi\}{\rm tr}\{\tilde{G},g\}+
\frac{1}{2}{\rm tr}\{f,\tilde{G}\}{\rm tr}\{\phi,g\}+\\
 & & \frac{1}{2}{\rm tr}\{f,\phi\}{\rm tr}(\chi\{\phi,g\})-
\frac{1}{2}{\rm tr}(\{f,\phi\}\chi){\rm
tr}\{\phi,g\}]\,d^{3}x,\nonumber\\
 & & \phi^{\alpha\beta}\stackrel{\rm def}{=}\pi^{\alpha}*\pi^{\beta}
\nonumber
\end{eqnarray}

\noindent Procedure of performing trace refers to indices of
functions
$\tilde{G},\phi,\chi$, while integration variable $x$ is their
argument.
Dirac bracket turns out to be nonpolinomial (due to occurence of
$(\det\|
g_{\alpha\beta}\|)^{-2}$). Also note that different components of
$\omega$
do not commute.

\bigskip
4. Finally, let us pass to Ashtekar variables and decompose for that
antisymmetric tensors $A^{ab}$ into selfdual $\,^{+}\!A$ and
antiselfdual
$\,^{-}\!A$ parts,
\begin{equation}
A=\,^{+}\!A+\,^{-}\!A,~~~^{\pm}\!A=\frac{1}{2}(A\pm i\,^{*}\!A),
{}~~~i\,^{*}\!(\,^{\pm}\!A)=\pm(\,^{\pm}\!A),
\end{equation}

\noindent each of which embed into complex 3D vector space by
expanding
over basis of (anti-)selfdual matrices
\begin{equation}
^{\pm}\!\Sigma^{k}_{ab}=\pm i(\delta^{k}_{a}\delta^{0}_{b}-
\delta^{k}_{b}\delta^{0}_{a})+\epsilon_{kab},
\end{equation}

\noindent so that
\begin{equation}
^{\pm}\!A^{ab}=\,^{\pm}\!A^{k}\,^{\pm}\!\Sigma^{ab}_{k}/2
\stackrel{\rm def}{=}\,^{\pm}\!\vec{A}\cdot\,^{\pm}\!\vec{\Sigma}
^{ab}/2
\end{equation}

\noindent (matrices $-i\,^{\pm}\!\Sigma^{a}_{kb}$ are chosen to
obey algebra of Pauli matrices $\sigma^{k}$). Then for real tensor
quantity $\,^{+}\!\vec{A}=\overline{\,^{-}\!\vec{A}}$ (overlining
means
usual complex conjugation).

At such embedding the constraints become sums or differences between
monoms of only selfdual and of only antiselfdual fieds. It is
convenient
to group these as follows:
\begin{eqnarray}
\label{pi-cdot-pi}
2i\phi^{\alpha\beta} & = & \,^{+}\!\vec{\pi}^{\alpha}\cdot\,^{+}\!
\vec{\pi}^{\beta}-\,^{-}\!\vec{\pi}^{\alpha}\cdot\,^{-}\!\vec{\pi}
^{\beta}=0 \\
\label{pi-pi-cdot-D-pi}
-2i\tilde{G}^{\alpha\beta} & = & \,^{+}\!\vec{\pi}^{\gamma}\cdot
\,^{+}\!\vec{\pi}^{(\alpha}\times\,^{+}\!{\cal D}_{\gamma}
\,^{+}\!\vec{\pi}^{\beta)}-\,^{-}\!\vec{\pi}^{\gamma}\cdot\,^{-}\!
\vec{\pi}^{(\alpha}\times\,^{-}\!{\cal D}_{\gamma}\,^{-}\!\vec{\pi}^
{\beta)}=0 \\
\label{D-pi-minus-D-pi}
C\circ (\,^{+}\!\vec{\Sigma}-\,^{-}\!\vec{\Sigma})/2 & = & \,^{+}\!
{\cal D}_{\alpha}\,^{+}\!\vec{\pi}^{\alpha}-\,^{-}\!{\cal
D}_{\alpha}
\,^{-}\!\vec{\pi}^{\alpha}=0 \\
\label{D-pi-plus-D-pi}
C\circ (\,^{+}\!\vec{\Sigma}+\,^{-}\!\vec{\Sigma})/2 &
= & \,^{+}\!{\cal D}_{\alpha}\,^{+}\!\vec{\pi}^{\alpha}+
\,^{-}\!{\cal D}_{\alpha}\,^{-}\!\vec{\pi}^{\alpha}=0 \\
\label{pi-cdot-R}
2{\cal H}_{\alpha} & = &
\epsilon_{\alpha\beta\gamma}(\,^{+}\!\vec{\pi} 
^{\beta}\cdot\,^{+}\!\vec{R}^{\gamma}+\,^{-}\!\vec{\pi}^{\beta}
\cdot\,^{-}\!\vec{R}^{\gamma})=0 \\
\label{pi-pi-cdot-R}
-4{\cal H}_{0} & = & \epsilon_{\alpha\beta\gamma}(\,^{+}\!
\vec{\pi}^{\alpha}\times\,^{+}\!\vec{\pi}^{\beta}\cdot\,^{+}\!
\vec{R}^{\gamma}+\,^{-}\!\vec{\pi}^{\alpha}\times\,^{-}\!
\vec{\pi}^{\beta}\cdot\,^{-}\!\vec{R}^{\gamma})=0.
\end{eqnarray}

\noindent Here $\,^{\pm}\!{\cal D}_{\alpha}(\cdot)=\partial_{\alpha}
(\cdot)-\,^{\pm}\!\vec{\omega}_{\alpha}\times(\cdot), \,^{\pm}\!
\vec{R}^{\alpha}=-\epsilon^{\alpha\beta\gamma}[\,^{\pm}\!{\cal D}_
{\beta},\,^{\pm}\!{\cal D}_{\gamma}]/2$, while $(\alpha\dots\beta)$
means the sum of objects with indices $\alpha\dots\beta$ and $\beta
\dots\alpha$. Equations (\ref{pi-cdot-pi}), (\ref{pi-pi-cdot-D-pi})
and (\ref{D-pi-minus-D-pi}) at $\,^{+}\!\vec{\pi}=\overline{\,^{-}\!
\vec{\pi}}$ present $6+6+3$ real conditions. If the condition
$\,^{+}\!\vec{\pi}=\overline{\,^{-}\!\vec{\pi}}$ is not assumed
beforehand, we deal with $6+6+3$ {\bf complex} equations on $\,^{+}
\!\vec{\pi}, \,^{+}\!\vec{\omega}$. It follows from
(\ref{pi-cdot-pi})
that some $U$ exists, an element of $SO(3,C)$, such that
$\,^{+}\!\vec{\pi}
^{\alpha}=U\,^{-}\!\vec{\pi}^{\alpha}$. Then (\ref{pi-pi-cdot-D-pi})
and (\ref{D-pi-minus-D-pi}) allow us to find 9 components of
connection:
$\,^{+}\!{\cal D}_{\alpha}=U\,^{-}\!{\cal D}_{\alpha}U^{\dag}$.
As a result, (\ref{D-pi-plus-D-pi}) - (\ref{pi-pi-cdot-R}) are
fullfilled
separately for $(+)$ and $(-)$-components. Thus we arrive at
the Ashtekar constraints:
\begin{equation}
{\cal D}_{\alpha}\vec{\pi}^{\alpha},~~~\epsilon_{\alpha\beta\gamma}
\vec{\pi}^{\beta}\cdot\vec{R}^{\gamma},~~~\epsilon_{\alpha\beta\gamma}
\vec{\pi}^{\alpha}\times\vec{\pi}^{\beta}\cdot\vec{R}^{\gamma},
\end{equation}

\noindent where $\vec{\pi},\vec{R}$ are $(+)$ or $(-)$-components.

Equations (\ref{pi-cdot-pi}), (\ref{pi-pi-cdot-D-pi}) in the theory
with
pseudoEuclidean signature are known as reality conditions
\cite{real}.
These equations, however, survive in the real theory with Euclidean
signature. In both cases these connect self- and antiselfdual sectors
of
theory and their being the II class constraints leads to commutators
(Dirac brackets) of $\pi,\omega$ different from canonical ones and
nonpolinomial (see (\ref{Dir-brack})). In particular, different
components
of $\omega$ do not commute and there is no such representation of
commutation relations that $\omega$ be $c$-number. On the other
hand,
the components of $\pi$ do not commute with each other, and one might
try
to use representation in which $\pi$ is $c$-number. However, when
imposing
the I class constraints on states we shall not get analytical
functionals
of $\pi$ as solutions (physical case of pseudoEuclidean signature is
considered). Nonanalyticity occurs in the dependence of commutators
on
the (real) metric $g_{\alpha\beta}$. Indeed, in selfdual components
$g^{\alpha\beta}\!\det\|g_{\gamma\delta}\|=\pi^{\alpha}\circ~\pi^{\beta}$
becomes {\bf real part} of
$2\vec{\pi}^{\alpha}\cdot~\vec{\pi}^{\beta}$.

Besides, self- and antiselfdual components do not commute. For
example,
it is easy to find from general formula (\ref{Dir-brack}) that
\begin{equation}
\{\,^{+}\!\omega^{i}_{\alpha}(x),\,^{-}\!\pi^{\beta k}(x^{\prime}\}
_{\rm D}=-(\det\|g_{\alpha\beta}\|)^{-1}(\,^{+}\!\pi^{\gamma i}
\,^{-}\!\pi_{\gamma k}\delta^{\beta}_{\alpha}+\,^{+}\!\pi^{\beta i}
\,^{-}\!\pi_{\alpha k})\delta^{3}(x-x^{\prime}).
\end{equation}

\bigskip
We thus have considered 2 formulations of GR with polinomial
Lagrangian.
These are defined by the sets of constraints
(\ref{I-class}), (\ref{II-class}) and (\ref{I-class-2}),
(\ref{pi-omega-II-class}) on phase spaces of pairs $(\pi,\omega),
(\tilde{n},n)$ and of $(\pi,\omega)$, respectively. The first
version
turns out to be appropriate for generalisation to discrete Regge
gravity
where it allows one to put theory into quasipolinomial form
\cite{Kha}.
Second version leads, on separating self- and antiselfdual components
of
tensors, to Ashtekar variables. It turns out that due to commutators
Ashtekar variables are not free from nonpolinomiality and
nonanalyticity.


\newpage
\begin{thebibliography}{99}
\bibitem{Ash}
 Ashtekar, A.(1986) {\it Phys.Rev.Lett.}, {\bf 57}, 2224.\\
 Ashtekar, A.(1987) {\it Phys.Rev.}, {\bf D36}, 1787.
\bibitem{Jac}
 Jacobson, T., and Smolin, L.(1988) {\it Nucl.Phys.}, {\bf B299},
295.
\bibitem{Schw}
 Schwinger, J.(1963) {\it Phys.Rev.}, {\bf 130}, 1253.
\bibitem{real}
 Ashtekar, A., Romano, J.D., and Tate, R.S.(1988) "New variables for
 gravity: inclusion of matter" Syracuse University preprint.
\bibitem{Kha}
 Khatsymovsky, V."Regge calculus in the canonical form",
 in press.
\end{thebibliography}
\end{document}